\documentclass[9pt,twocolumn,twoside]{optica}
\setboolean{shortarticle}{true}
\setboolean{minireview}{false}

\usepackage{siunitx}
\usepackage{lmodern}
\usepackage{xspace}

\renewcommand\textemdash{\leavevmode\unskip\kern0.8pt\rule[0.19\baselineskip]{8pt}{0.4pt}\kern1pt\ignorespaces}

\hypersetup{
 pdftitle = {A high numerical aperture (NA = 0.92) objective lens for imaging and addressing of cold atoms},
 pdfauthor = {Carsten Robens, Stefan Brakhane, Wolfgang Alt, Felix Kleißler, Dieter Meschede, Geol Moon, Gautam Ramola, and Andrea Alberti}
}

\pdfoutput=1

\definecolor{wildstrawberry}{rgb}{1.0, 0.26, 0.64}
\definecolor{mediumorchid}{rgb}{0.73, 0.33, 0.83}
\definecolor{limegreen}{rgb}{0.2, 0.8, 0.2}
\definecolor{fluorescentorange}{rgb}{1.0, 0.75, 0.0}
\definecolor{egyptianblue}{rgb}{0.06, 0.2, 0.65}

\newcommand{\figref}[2]{\hyperref[#1]{\getrefnumber{#1}(#2)}}

\newif\ifarXiv
\arXivtrue
\arXivtrue

\let\oldsection\section

\newcommand{\test}[1]{{#1}.\textemdash}

\makeatletter
\renewcommand\section{\@startsection {section}{1}{\z@}%
                                   {-0pt \@plus +4pt \@minus 0pt}%
                                   {0em}%
                                   {\sffamily\bfseries\raggedright\test}}
\DeclareRobustCommand{\@seccntformat}[1]{%
  \def\temp@@a{#1}%
  \def\temp@@b{section}%
  \ifx\temp@@a\temp@@b
  \csname the#1\endcsname .\quad%
  \else
  \csname the#1\endcsname\quad%
  \fi
}
\makeatother

\ifarXiv
\makeatletter
\let\old@maketitle\@maketitle
\renewcommand{\@maketitle}{%
\vskip-68pt%
\old@maketitle%
}
\makeatother

\makeatletter
\renewcommand{\abscontent}{
\noindent
{%
\parbox{\dimexpr\linewidth}{%
	\absfont \theabstract
 \copyrightfont \ifthenelse{\boolean{displaycopyright}}{}{}
}%
}%
\vskip8pt%
\noindent
\hfil\rule{\linewidth}{.4pt}
}
\makeatother
\else
\makeatletter
\let\old@maketitle\@maketitle
\def\@maketitle{\vskip-25pt\old@maketitle}
\makeatother
\fi

\title{A high numerical aperture (NA = 0.92) objective lens for imaging and addressing of cold atoms}

\author[1,*]{Carsten Robens}
\author[1]{Stefan Brakhane}
\author[1]{Wolfgang Alt}
\author[1]{Felix Kleißler}
\author[1]{Dieter Meschede}
\author[1]{Geol Moon}
\author[1]{Gautam Ramola}
\author[1]{Andrea Alberti}

\affil[1]{Institut f\"ur Angewandte Physik, Universit\"at Bonn, Wegelerstr.~8, D-53115 Bonn, Germany}

\affil[*]{Corresponding author: robens@iap.uni-bonn.de}

\dates{Compiled \today}

\ociscodes{(020.0020) Atomic and molecular physics; (110.0110) Imaging systems; (120.3620) Lens system design; (180.0180) Microscopy; (270.0270) Quantum optics.}

\doi{\url{http://dx.doi.org/10.1364/optica.XX.XXXXXX}}

\begin{abstract}
	We have designed, built, and characterized a high-resolution objective lens that is compatible with an ultra-high vacuum environment.
	The lens system exploits the principle of the Weierstrass-sphere solid immersion lens to reach a numerical aperture (NA) of 0.92.
	Tailored to the requirements of optical lattice experiments, the objective lens features a relatively long working distance of \SI{150}{\micro\meter}.
	Our two-lens design is remarkably insensitive to mechanical tolerances in spite of the large NA.
	Additionally, we demonstrate the application of a tapered optical fiber tip, as used in scanning near-field optical microscopy, to measure the point spread function of a high NA optical system.
	From the point spread function, we infer the  wavefront aberration for the entire field of view of about \SI{75}{\micro\meter}.
	Pushing the NA of an optical system to its ultimate limit enables novel applications in quantum technologies such as quantum control of atoms in  optical microtraps with an unprecedented spatial resolution and photon collection efficiency.
\end{abstract}

\setboolean{displaycopyright}{true}

\begin{document}

\maketitle
\thispagestyle{fancy}
\ifthenelse{\boolean{shortarticle}}{\abscontent}{}

\section*{Introduction}
\label{sec:Introduction}
	High-resolution in-situ imaging of individual atoms in optical lattices has become an indispensable technique in modern quantum optics experiments~\cite{Ott:2016}, since it offers the capability to study quantum effects at the single particle level.
	This has been used, for example, to directly observe bosonic~\cite{Bakr:2009bxa,Sherson:2010hg} and fermionic Mott insulator states~\cite{Haller:2015hi,Cheuk:2015jr,Omran:2015io,Parsons:2016gr}.
	In addition, objective lenses with high NA are essential components to optically control the quantum state of individual atoms trapped in optical lattice with focussed laser beams~\cite{Weitenberg:2011gn,Wang:2015dx,Xia:2015eda,Wang:2016ev}, and can find application for the simulation of topological phases \cite{Groh:2016ff}.
	For these applications an improvement in the resolution directly leads to higher achievable gate fidelities.
	The recent success of high-resolution single-atom imaging and addressing is based on (a) technological advances towards objective lenses with higher NAs~\cite{Bakr:2009bxa,Sherson:2010hg} and (b) numerical super-resolution algorithms, capable of retrieving the position of individual atoms far beyond the optical resolution~\cite{Alberti:2016iy}.
	Recently, experimental realizations achieved NAs ranging from 0.68~\cite{Sherson:2010hg,Haller:2015hi}, to 0.75~\cite{Yamamoto:2016cy}, up to 0.87~\cite{Parsons:2015ck}, which rely on complex multi-lens systems.
	 	Ultimately, the optical resolution of an imaging system for ultracold atoms trapped in a vacuum environment is bound by $\lambda/2$ (Abbe criterium for NA=1) when photon collection reaches a solid angle of $2\pi$.

\begin{figure}[tb]
		\centering
		\fbox{\includegraphics[width=\linewidth]{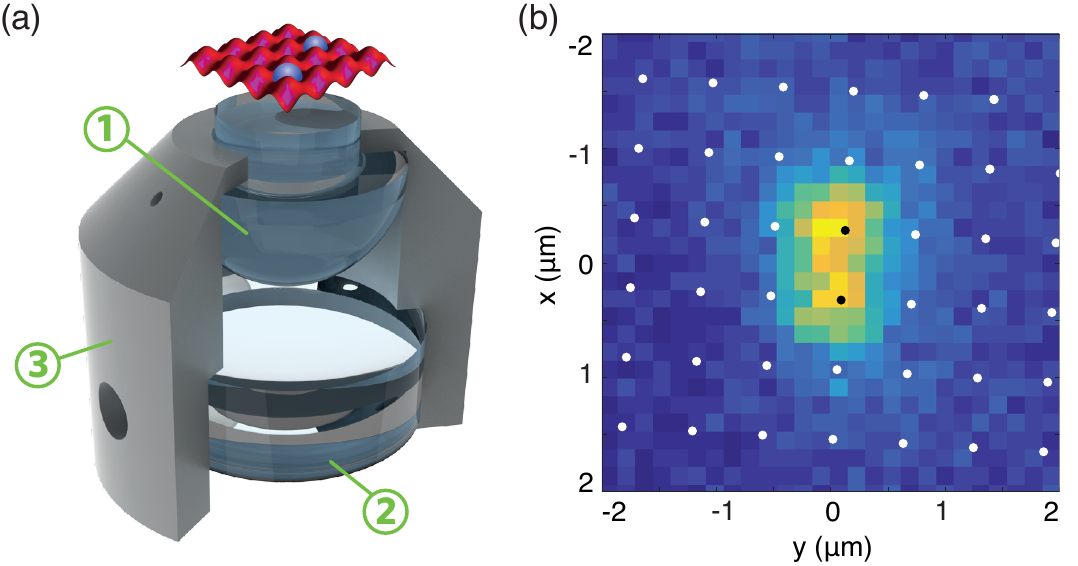}}
		\caption{\textbf{(a)} Cutaway drawing of the objective lens: (1) Weierstrass-like lens, (2) aspheric lens, (3) ceramic holder. 
		For illustrative purposes, atoms in the optical lattice in front of the objective lens are shown not to scale.
		\textbf{(b)} Fluorescence image acquired with our high NA objective lens, showing two cesium atoms trapped in adjacent lattice sites of a two-dimensional optical lattice.
		The positions of the optical lattice sites are indicated by white dots whereas the black dots correspond to the reconstructed positions of the two atoms.}
		\label{fig:ObjectiveAndAtomsInLattice}\vspace{-4mm}
	\end{figure}

	In this Letter, we present a novel lens design specifically tailored to the requirements of ultracold atom experiments which, using only two lenses, achieves diffraction-limited imaging under ultra-high vacuum conditions with an unprecedented NA of \num{0.92}.
	Optimized for \SI{852}{\nano\meter}, the objective lens depicted in Fig.~\figref{fig:ObjectiveAndAtomsInLattice}{a} has a resolution of \SI{460}{\nano\meter} (Abbe criterion) and a field of view spanning over \SI{75}{\micro\meter}, which corresponds to more than $100\times100$ lattice sites in a typical optical lattice.
	Figure~\figref{fig:ObjectiveAndAtomsInLattice}{b} shows a close-up, recorded with our objective lens, of two adjacent cesium atoms, which are trapped in an optical lattice with a lattice constant of \SI{612}{\nano\meter}.

\section*{Lens design}
\label{sec:LensDesign}
	Due to their large collection angle, objective lenses with very high NA (above 0.9) have a short working distance \textemdash the distance between the last surface and the focal plane \textemdash which generally does not exceed \SI{1}{\milli\meter}, even if multiple lens elements are employed.
	Consequently, in order to image ultracold atoms suspended in an optical trap, such an objective lens must be placed directly inside the vacuum chamber.
	However, this is not feasible using commercially available objective lenses with high NA, as these are not compatible with an ultra-high vacuum environment. In fact, in order to align multiple lens elements within the specified mechanical tolerance, commercial objective lenses often employ soft structural adhesives \cite{YoderJr:2015} that have too high outgassing rates and cannot withstand a high-temperature bake-out, which is required to achieve an ultra-high vacuum.
	Therefore, most ultracold-atom microscopes \cite{Alt:2002fya,Yamamoto:2016cy,Sherson:2010hg} have traded a very high numerical aperture for a long working distance, so that the entire objective lens can be placed outside the vacuum chamber.
	In this approach, the vacuum window must be taken into account in the design of the lens system to avoid optical aberrations.

	Recently, higher NAs have been achieved in ultracold-atom experiments by employing a hemispherical lens inside the vacuum chamber, coupled to a moderate NA objective lens outside~\cite{Bakr:2009bxa}.
	Similar to solid immersion lens microscopy~\cite{,Wu:2000cv}, the hemispherical lens of refraction index $n$ enhances the NA of the overall optical system by a factor $n$ compared to the NA of the coupling lens system situated outside of vacuum.
	Instead of a  hemispherical lens, our objective-lens design is based on a Weierstrass sphere\textemdash a truncated sphere with a thickness equal to $(1+1/n)$ times its radius.
	The advantage of this geometry is a larger NA enhancement factor of $n^2$ compared to  $n$, which greatly lessens the complexity of the subsequent coupling lens system\textemdash in our case, a single aspheric lens with $\mathrm{NA}=0.35$.{}
	Of course, the overall NA of such objective lenses operating in vacuum (or air) can not exceed 1, which is realized once the entire half solid angle is collected.
	Moreover, along with the NA enhancement, the diffraction-limited chromatic bandwidth is strongly reduced by the Weierstrass sphere \cite{Shimura:2000}; this however has no bearing in applications such as fluorescence imaging at well-defined wavelengths.

	Our objective lens is designed for fluorescence imaging of cesium atoms at the diffraction limit with a NA of 0.92.
	Such a NA corresponds to a large photon collection efficiency covering almost 1/3 of the full solid angle, which results in a resolution of \SI{460}{\nano\meter}.
	The	design of the objective lens has been optimized to increase its working distance\textemdash vanishing for a solid immersion objective lens \cite{Baba:1999}\textemdash to \SI{150}{\micro\meter}, which is sufficiently long to form a two-dimensional optical lattice of counter-propagating laser beams in the focal plane, as shown in Fig.~\figref{fig:ObjectiveAndAtomsInLattice}{a}.
	Due to the increased working distance, the Weierstrass-like lens introduces a significant amount of aberrations, which are compensated by carefully tailoring the surface profile of the aspheric lens up to the tenth-order correction to an axially symmetric quadric surface.
To that purpose, we used a self-developed scalar ray tracing software, which allows us to simultaneously optimize all relevant parameters.
	In addition, the design of our infinity-corrected objective lens does not depend on the vacuum window.
	The optical properties of the objective lens are summarized in Tab.~\ref{tab:MicroscopeObjectiveParameterComparision} and the details of the lens surfaces are given in the Supplementary Material. 
	Vectorial corrections to optical diffraction, not considered in our design, are expected to affect only marginally the imaging performance; however, they could have perceptible effects in applications of the objective lens such as coherent control of ultracold atoms \cite{Lodahl:2016}.
	Furthermore, to suppress stray light, all surfaces have an anti-reflective coating for wavelengths between $\SI{840}{nm}$ and $\SI{900}{nm}$.
	The planar front surface of the Weierstrass-like lens has an additional high-reflective coating for \SI{1064}{nm}, which allows us to form an additional optical lattice by retro-reflection (not shown in Fig.~\figref{fig:ObjectiveAndAtomsInLattice}{a}).

	We carried out a thorough tolerance analysis of our objective lens, taking into account
	 deviations from the design lens-surface profile, thickness, and refractive index, as well as misalignments of the lenses such as tilts ($\approx\SI{0.5}{\degree}$) and displacements ($\approx\SI{10}{\micro\meter}$) along the lateral and axial direction, respectively.
	All deviations were chosen to exceed manufacturing tolerances by at least one order of magnitude. 	The result of our tolerance analysis shows that our objective-lens design is remarkably insensitive to parameter deviations in spite of the large NA and,
	most importantly, that it remains diffraction-limited (Strehl ratio $>0.8$) even in the worst-case scenario of all deviations acting concurrently.

	Both lenses are made of N-SF10 glass from Schott and manufactured by Asphericon GmbH. Operating in a clean-room environment, we mounted the lenses inside a ceramic (Al$_2$O$_3$) holder manufactured by BeaTec GmbH using an ultra-high vacuum compatible adhesive (Epotek H77).
	Due to the insensitivity to lens tilts and displacements, our assembly procedure relies only on the manufacturing precision (a few $\si{\micro\meter}$) of the components, without need for a direct inspection of the objective lens' optical performance.
	The planar side of the Weierstrass-like lens (see Fig.~\figref{fig:ObjectiveAndAtomsInLattice}{a}) is particularly designed to include a mechanical stop, allowing the distance between the two lenses\textemdash one of the few critical parameters \textemdash to be precisely set and, furthermore, preventing any clipping of the optical-lattice laser beams. 
	We ensure ultra-high vacuum compatibility by adding venting holes in the ceramic holder and by matching the thermal expansion coefficients of the lenses and the holder to enable bake-out temperatures up to \SI{150}{\celsius}.

{\renewcommand{\arraystretch}{1.2}
\begin{table}[b]
\vspace{-5mm}
\centering
\caption{\label{tab:MicroscopeObjectiveParameterComparision}\textbf{Key properties of the design of the NA=0.92 objective lens.} Starred properties are defined by a reduction of the Strehl ratio to 0.8 (diffraction limit).}
\vspace*{1mm}
\begin{tabular}{lc}
	\hline
	Design wavelength $\lambda$ & \SI{852}{\nano\meter}\\
	Optical resolution $\left[\lambda/(2\hspace{1pt}\mathrm{NA})\right]$ & \SI{460}{\nano\meter}\\ 
	On-axis Strehl ratio & \SI{0.99}{}\\ 
	Collection angle & \SI{134}{\degree}\\
	Collection solid angle & $4\pi\times0.30$\,\si{\steradian}\\
	Working distance & \SI{150}{\micro\meter}\\
	Field of view$^{*}$ & $\pm$\SI{38}{\micro\meter}\\
	Effective focal length $f_\text{eff}$ & \SI{11.96}{\milli\meter}\\
	Depth of focus $\left[\lambda/(2\,\mathrm{NA^2})\right]$& $\pm\SI{250}{\nano\meter}$\\
	Chromatic bandwidth$^{*}$ & $\pm\SI{1.3}{\nano\meter}$\\
	Chromatic bandwidth with refocussing$^{*}$ & $\pm\SI{17}{\nano\meter}$\\
	Collimated beam diameter & \SI{22}{\milli \meter}\\
	\hline
\end{tabular}
\end{table}}

\section*{Optical characterization of the high NA objective lens}
	\begin{figure}[t]
		\centering
		\fbox{\includegraphics[width=\linewidth]{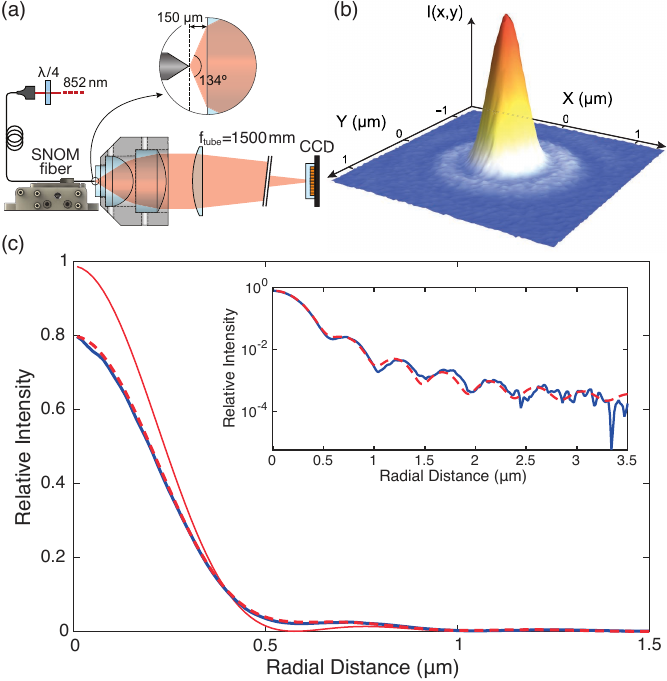}}
		\caption{\textbf{(a)} Optical setup employed to measure the point spread function. 
		The circularly polarized light emitted from the SNOM fiber tip is collimated through the high NA objective lens, and focused onto a beam-profiling CCD camera using a tube lens.
		\textbf{(b)} Two-dimensional point spread function recorded with the beam profile CCD-camera.
		\textbf{(c)} The azimuthally integrated point spread function for: measured data (solid blue line), fitted model based on a wavefront expansion in Zernike polynomials (dashed red line), the same as the latter but with defocus aberration set to zero (solid red line).
		The inset attests the quality of the fited model in a logarithmic scale.}
		\label{fig:SNOMSetupAndPSF}
		\ifarXiv
		\else
		\vspace{-5mm}
		\fi
	\end{figure}
	An effective characterization of the optical performance of an objective lens is  obtained by measuring its point spread function (PSF)\textemdash the image of a point-like emitter such as small particles or pin holes of a size much smaller  than than the optical resolution~\cite{Juskaitis:2006ic}.
	Commercially available pinholes ($\varnothing\approx\SI{1}{\micro\meter}$) represent a suitable solution for objective lenses with low to moderate NAs, while they cannot be applied for those with high NAs.
	We here suggest a different method suited for high NA objective lenses, which uses the light emitted from an aluminum-coated tapered optical fiber tip.
	Such fibers are typically employed for scanning near-field optical microscopy (SNOM)~\cite{Betzig:1993gd} and can be commercially obtained with tip diameters as small as \SI{100}{\nano\meter}.
	Compared to light-emitting nanoparticles, SNOM fiber tips have several advantages, including the preservation of arbitrary polarizations~\cite{Obermuller:1995kq}, a higher signal-to-noise ratio due to virtually zero ambient stray light, and a less demanding optical setup.
	To obtain the PSF, we image circularly polarized light emitted from a SNOM fiber tip (Lovalite E50-MONO780-AL-200) with the objective lens mounted in an infinity-corrected microscope configuration, as illustrated in Fig.~\figref{fig:SNOMSetupAndPSF}{a}; using circular polarization provides us directly with an averaged PSF for the two linear polarization components. 
	The SNOM fiber tip is mounted on a multi-axis translation stage to allow its position to be adjusted with sub-micrometer precision while simultaneously monitoring the PSF with the CCD camera.
	Figure~\figref{fig:SNOMSetupAndPSF}{b} shows the recorded image of the radiation pattern emitted from the SNOM fiber tip, the shape of which closely resembles the well-known Airy disk.
	The image constitutes a direct measurement of the PSF of the objective lens, since the effects of a finite CCD pixel size are made negligible \cite{Alberti:2016iy} by the large magnification factor ($f_\text{tube}/f_\text{eff}\approx 125$), projecting the PSF's profile onto several CCD pixels. %
	As shown in Fig.~\figref{fig:SNOMSetupAndPSF}{c}, the maximum value of the measured PSF is about $\num{0.8}$ times that of an ideal Airy disk for NA=\num{0.92} (Strehl ratio); such a large Strehl ratio demonstrates that the performance of the objective lens at full NA is diffraction limited. %

	To gain insight into the optical performance of the lens system, we use the mathematical relation between the PSF and the wavefront, which includes all information about optical aberrations:
	\begin{equation}
		\label{eqn:PSFmathematical}
		\mathrm{PSF} = \left|\mathcal{F}\left\{\hspace{-1pt} P(x,y)\,e^{i\,2\pi\hspace{0.3pt}R(x,y)}\right\}\right|^2\,,
	\end{equation}
	where $\mathcal{F}\{\cdot\}$ is the two-dimensional Fourier transformation, $P(x,y)$ is the real-valued pupil function depending on the illumination intensity in the back-focal plane, and $R$ is the wavefront, which can be expressed through a series expansion in Zernike polynomials~\cite{BornWolf}.
	For a homogeneous pupil function and vanishing aberrations (planar wavefront), \eqref{eqn:PSFmathematical} yields exactly the Airy disc.
	While the pupil function can be regarded as homogeneous for objective lenses with low and moderate numerical apertures, for high NA objective lenses additional effects of such as lens apodization and the azimuthal dependence of the radiation pattern emitted from the fiber tip must be taken into account.
	The details of these effects are discussed in the Supplementary Material.

	\begin{table*}[tb]
		\centering
		\caption{\bf Result of the wavefront fitting to the measured PSF expressed in terms of low-order Zernike polynomials.
		The overall wavefront distortion is obtained by adding the different contributions in quadrature.}
		\begin{tabular}{lcccccc}
			\hline
			& Defocus & Astigmatism & Coma & Trefoil & Spherical & Secondary astigmatism \\
			\hline
			Orders {(radial, azimuthal)} & (2,0) & (2,2) & (3,1) & (3,3) & (4,0) & (4,2)\\
			RMS wavefront {distortion ($\lambda$ units)} & $0.07(3)$ & $-0.010(1)$ & $-0.004(1)$ & $-0.001(1)$ & $0.006(1)$ & $-0.015(1)$\\
			\hline
		\end{tabular}
		\label{tab:ZernikePSF}
	\end{table*}

	To reconstruct the optical aberrations affecting the objective lens, we fitted by non-linear least squares minimization the PSF computed from \eqref{eqn:PSFmathematical} using an expansion in the lowest-order Zernike polynomials directly to the measured intensity distribution displayed in Fig.~\figref{fig:SNOMSetupAndPSF}{b}.
	The results are shown in Fig.~\figref{fig:SNOMSetupAndPSF}{c}.
	The fitted model confirms a diffraction-limited performance of the objective lens and, moreover, reveals a numerical aperture of $\mathrm{NA}={0.938}\pm{0.001}$, which exceeds the design value.
	We attribute the difference of the numerical aperture to an uncertainty in the radius of the hard aperture stop ($\approx\SI{200}{\micro\meter}$) located near the back-focal plane of the objective lens in our characterization setup. 
	The reconstructed Zernike coefficients are given in Tab.~\ref{tab:ZernikePSF}, demonstrating that the wavefront aberration is dominated by the defocus contribution ($\lambda/14$).
	We attribute its relatively large value to the limited positioning precision ($\approx\SI{200}{\nano\meter}$) of the employed translation stage.
	\begin{figure}[b!]
		\centering
		\fbox{\includegraphics[width=\linewidth]{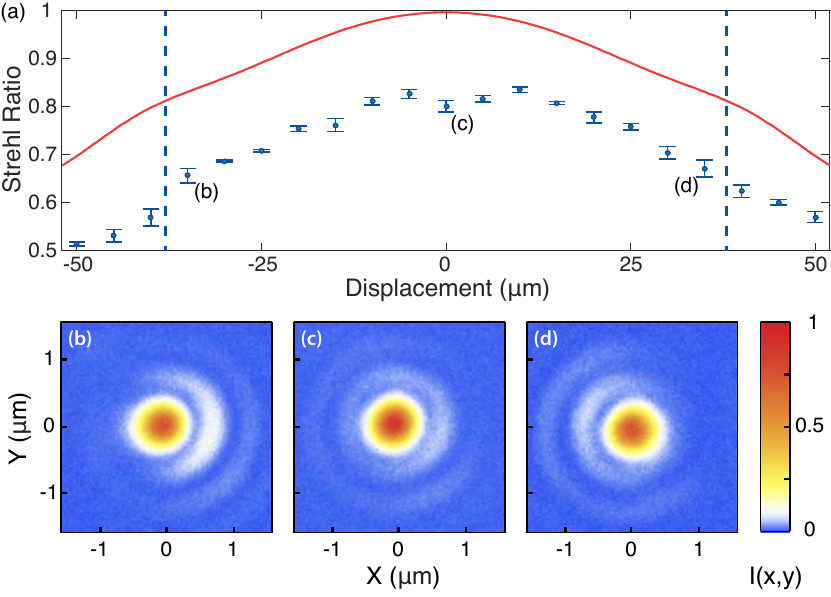}}
		\caption{\textbf{(a)} Measured values (blue dots) and		
		theoretical prediction based on the objective lens design (red line) as a function of the transversally displaced point-like emitter.
		The region between the vertical dashed lines represents the expected field of view ($\pm\SI{38}{\micro\meter}$) of the objective lens, in which the design Strehl ratio is above \num{0.8}.
		\textbf{(b)}-\textbf{(d)} Example images of the corresponding intensity distributions at different positions of the field of view.
		}
		\label{fig:FoVMeasurement}
		\vspace{-3mm}
	\end{figure}
	We determine the extent of the field of view by repeating the foregoing analysis for increasing transverse displacements of the SNOM fiber tip.
	The results displayed in Fig.~\ref{fig:FoVMeasurement} show that
	the recorded Strehl ratios agree well with the theoretical prediction based on the objective lens design, except for a systematic downward shift caused by the already-mentioned defocus.

\section*{Conclusions and outlook}
	In this Letter we presented a diffraction-limited infinity-corrected high NA objective lens based on a novel design requiring only two lenses, which enable it to be mounted inside an ultra-high vacuum chamber.
	While our objective lens is specifically tailored to the requirements of experiments with ultracold atoms,
	it could also find application in other quantum-optics experiments, especially in those employing cryogenic temperatures such as vacancy centers, quantum dots, and cryogenic surface ion traps.
	The design itself can be readily adapted to other wavelengths  as well.
	By the way of example, we provide design information in the Supplementary Material (through OSLO computer files) for the wavelengths \SI{852}{\nano\meter} (suited for Cs), \SI{780}{\nano\meter} (suited for Rb), \SI{671}{\nano\meter} (suited for Li), \SI{461}{\nano\meter} (suited for Sr), and \SI{399}{\nano\meter} (suited for Yb). Our self-developed ray tracing software is made available upon request.

\section*{Acknowledgments}
We are indebted to U.~Schomaecker for independently verifying our lens design.
We thank E.~Roth from the Carl Zeiss Jena GmbH for valuable discussions during the early stage of the design process.
We also thank the Center of Advanced European Studies and Research (CAESAR) for providing access to their clean room facilities.
We acknowledge financial support from the NRW-Nachwuchsforschergruppe ``Quantenkontrolle auf der Nanoskala'', and the ERC grant DQSIM.
C.R.\ acknowledges support from the Studienstiftung des deutschen Volkes, and C.R., S.B., and G.R.~from the Bonn-Cologne Graduate School.

\let\section\oldsection

\ifarXiv
\makeatletter
\expandafter\protected@edef\csname refname\@suffix\endcsname{References}%
\makeatother

\bibliography{references}

\else
\vspace{1mm}\noindent
\href{http://www.opticsinfobase.org/submit/style/supplementary-materials-optica.cfm}{See Supplement 1 for supporting content.}

\makeatletter
\let\old@lbibitem\@lbibitem
\def\@lbibitem[#1]#2{\old@lbibitem[#1]{new#2}}
\makeatother

\bibliography{references}

\makeatletter
\def\@lbibitem[#1]#2{\old@lbibitem[#1]{#2}}
\makeatother

\ifthenelse{\boolean{shortarticle}}{%
\clearpage
\bibliographyfullrefs{references}
}{}
\fi

\end{document}